\documentclass[twocolumn,showpacs,preprintnumbers,amsmath,amssymb,floatfix,nofootinbib]{revtex4-1}

\usepackage{amsmath,amsfonts,bbm,euscript,hyperref}
\usepackage{graphicx}
\usepackage{dcolumn}
\usepackage{multirow}
\usepackage{float}
\usepackage{booktabs,makecell}
\usepackage{bm}
\usepackage{epsfig}
\epsfclipon
\usepackage{enumitem}

\usepackage{balance}

\newcommand{\nco}{\newcommand}
\nco{\beq}{\begin{equation}} \nco{\eeq}{\end{equation}}
\nco{\beqa}{\begin{eqnarray}} \nco{\eeqa}{\end{eqnarray}}
\def\be{\begin{equation}}
\def\ee{\end{equation}}
\def\baray{\begin{eqnarray}}
\def\earay{\end{eqnarray}}

\nco{\lra}{\leftrightarrow}

\nco{\sss}{\scriptscriptstyle} \nco{\dphi}{\varphi}
\nco{\lsim}{\mbox{\raisebox{-.6ex}{~$\stackrel{<}{\sim}$~}}}
\nco{\gsim}{\mbox{\raisebox{-.6ex}{~$\stackrel{>}{\sim}$~}}}

\def\IK{\relax{\rm I\kern-.20em K}}
\def\IM{\relax{\rm I\kern-.20em M}}

\def\lsim{\mbox{\raisebox{-.6ex}{~$\stackrel{<}{\sim}$~}}}
\def\gsim{\mbox{\raisebox{-.6ex}{~$\stackrel{>}{\sim}$~}}}
\def\sss{\scriptscriptstyle}

\begin{document}

\preprint{UMN-TH-3906/19}

\title{Light scalar modes from holographic deformations}

\author{Yusuf Buyukdag}
\email{ysfbykdg@gmail.com}

\affiliation{School of Physics and Astronomy, University of Minnesota, Minneapolis, Minnesota 55455, USA}

\begin{abstract}
We consider different ways of modifying the mass spectrum of a strongly coupled gauge theory with confinement using the AdS/CFT correspondence. Single- and multitrace deformations are introduced, such that the resulting theory has a mode lighter than the confinement scale. The multitrace deformation is shown to be a possible way of achieving a light composite mode, unlike the single-trace deformation where the light mode is an admixture of elementary and composite states.
\end{abstract}

\maketitle

\section{Introduction}
\vspace{-4mm}
Based on our experience with QCD, we know that the composite states of a strongly coupled gauge theory have masses around the confinement scale. Usually, exceptions to this occur in the presence of an underlying symmetry as in the case of pions. Spontaneously broken global symmetries might be introduced to have light states when the symmetry is also broken explicitly. There have been previous studies that argued the existence of a light scalar from a five-dimensional perspective \cite{BitaghsirFadafan:2018efw, Pomarol:2019aae, Elander:2010wd}. In this paper, we explore alternative four-dimensional dual models for producing scalars lighter than the confinement scale of the theory in a simpler five-dimensional setup. 

One can modify the theory with no light mode in different ways such as coupling an elementary scalar to a composite operator or introducing effective interactions in the strong sector. These changes modify the mass spectrum of the theory; however, it is not straightforward to compute the effect due to the strong dynamics. Therefore, inspired by the AdS/CFT correspondence \cite{Maldacena:1997re}, we can construct the dual model in the extra-dimensional space where the modifications of the mass spectrum are more tractable. What is needed to have a light mode in the extra-dimensional model is straightforward, so we try to understand it from the four-dimensional perspective. In Sec.\ref{sec:review}, we present the five-dimensional model dual to a composite operator of a strongly coupled gauge theory with confinement. In Sec.\ref{subsec:linear mixing}, we review the traditional dictionary, where an elementary scalar is added to modify the mass spectrum, and we discuss an alternative way of introducing an elementary scalar for the same purpose. In Sec.\ref{subsec:multi trace}, we analyze the implications of an alternative dictionary that does not include an elementary scalar. In Sec.\ref{sec:relationship}, we explore the relationship between two dictionaries.

\section{Holography For a Massless Mode}\label{sec:review}

\noindent Consider the action for a real scalar field $\Phi (x, y)$, propagating on the five-dimensional anti-de Sitter (AdS$_5$) background
\begin{equation}
\vspace{1mm}
\mathcal{S} = \int d^5 x \sqrt{-g} \left[ - \frac{1}{2} \left( \partial_M \Phi \right)^2 - \frac{1}{2} ak^2 \Phi^2 \right],
\vspace{1mm}
\label{eq:fullAdS}
\end{equation}
where $a \geq -4$ parametrizes the bulk mass and $y$ is the coordinate of the fifth dimension. The metric for this background is 
\begin{equation}
\vspace{1mm}
ds^2 = e^{-2ky} \eta_{\mu \nu} dx^{\mu} dx^{\nu} + dy^2,
\vspace{1mm}
\end{equation}
where $k$ is the AdS curvature scale. Using the AdS/CFT dictionary, we know that this five-dimensional field corresponds to an operator $\mathcal{O}$, of dimension $\Delta$ (to be calculated later) in the four-dimensional theory \cite{Witten:1998qj}. The equation of motion derived from the variation of the action is 
\begin{equation}
\vspace{1mm}
\Box \Phi + e^{2ky} \partial_y \left( e^{-4ky} \partial_y \Phi \right) - ak^2 e^{-2ky} \Phi = 0,
\vspace{1mm}
\label{eq:eom}
\end{equation}
where $\Box = \eta^{\mu \nu} \partial_{\mu} \partial_{\nu}$. The solution for $\Phi$ can be written in momentum space as 
\begin{equation}
\vspace{1mm}
\Phi (p, y) = C_1 (p) e^{2ky} \left[ J_{\alpha} \left( \frac{ip}{k_y} \right) + C_2 (p) Y_{\alpha} \left( \frac{ip}{k_y} \right) \right], 
\vspace{1mm}
\label{eq:momentumspace}
\end{equation}
where $k_y \equiv ke^{-ky}$ and $\alpha \equiv \sqrt{4 + a}$. $C_1 (p)$ and $C_2 (p)$ are unknown functions that can be calculated once the boundary conditions are imposed. The solution is presented in this specific form just to be consistent with the solutions which will be presented later. It behaves near the AdS boundary $y \rightarrow - \infty$, as 
\begin{align}
\vspace{1mm}
\Phi (p, y) \rightarrow &- C_1 (p) C_2 (p) \frac{2^{\alpha}\Gamma (\alpha)}{\pi} \left( \frac{ip}{k} \right)^{-\alpha} e^{(2-\alpha)ky} \nonumber \\
&+ C_1 (p) \frac{2^{-\alpha}}{\Gamma (1 + \alpha)} \left( \frac{ip}{k} \right)^{\alpha} e^{(2+\alpha) ky}, \
\vspace{1mm}
\label{eq:nearUV}
\end{align}
which is consistent with Ref. \cite{Witten:1998qj}. Now, let us consider a slice of AdS$_5$, where the extra coordinate is compactified on an $S^1/\mathbb{Z}_2$ orbifold with UV and IR branes that exist at the orbifold fixed points $y_0$ and $y_1$, respectively \cite{Randall:1999ee}. It is customary to define a variable for the UV boundary value of the bulk field as $\hat{\Phi} (p) \equiv \Phi (p, y_0)$. The solution then becomes
\begin{equation}
\vspace{1mm}
\Phi (p, y) = \hat{\Phi} (p) e^{2k(y-y_0)} \frac{J_{\alpha} (ip / k_y) + C_2 (p) Y_{\alpha} (ip / k_y)}{J_{\alpha} (ip / k_{y_0}) + C_2 (p) Y_{\alpha} (ip / k_{y_0})},
\vspace{1mm}
\label{eq:bulksol}
\end{equation}
which minimizes the action in the bulk. Note that defining a variable for the UV boundary value does not specify the UV boundary condition yet. Minimizing the action on the branes as well forces one to satisfy
\begin{equation}
\vspace{1mm}
\left[ \left( \delta \Phi \right) \partial_y \Phi ~ \right]_{y=y_0, y_1} = 0,
\vspace{1mm}
\end{equation}
and boundary conditions depend on whether the boundary value is fixed, $\left[ \delta \Phi \right]_{y=y_0,y_1} = 0$, or not. Usually, the IR boundary value of the bulk field is not fixed, so the IR boundary condition is determined using $\left[ \delta \Phi \right]_{y=y_1} \neq 0$. The existence of an IR brane implies that the conformal symmetry is broken in the four-dimensional dual theory. While there was no mass scale in the conformal field theory (CFT) before introducing the UV and IR branes, now the composite states appear at the IR scale. The mass spectrum of the theory can be computed from the AdS/CFT correspondence. It is known that there is no massless mode for the gauge theory described by Eq.(\ref{eq:fullAdS}) in a slice of AdS$_5$ \cite{Gherghetta:2010cj}. The massless mode requires us to add brane masses in the following way,
\begin{align}
\vspace{1mm}
\mathcal{S} = \int d^5 x \sqrt{-g} \Bigg[ &- \frac{1}{2} \left( \partial_M \Phi \right)^2 - \frac{1}{2} ak^2 \Phi^2  \nonumber \\
				 	&- bk\Phi^2 \left[ \delta (y - y_0) - \delta (y - y_1) \right] \Bigg] , \nonumber \\
\vspace{1mm}
\label{eq:slicedAdS}
\end{align}
where $b \equiv 2 \pm \alpha$. The parameter range $b <2~(b > 2)$ is called the $-~(+)$ branch. Supersymmetry forces both brane masses to be related to each other in this form, so it can be generalized to different brane masses if the theory is not supersymmetric \cite{Gherghetta:2000qt}. These terms on the branes are necessary to have a massless mode, so we will explore what they imply in the four-dimensional dual theory later. The boundary equations are modified to
\begin{equation}
\vspace{1mm}
\left[ \left( \delta \Phi \right) \left( \partial_y - bk \right) \Phi ~ \right]_{y=y_0, y_1} = 0.
\vspace{1mm}
\end{equation}
The term on the IR brane simply modifies the IR boundary condition. As mentioned earlier, since we do not fix the IR boundary value, the IR boundary condition becomes $\left[ \left( \partial_y - bk \right) \Phi \right]_{y = y_1} = 0$. Plugging the solution (\ref{eq:bulksol}), with this IR boundary condition back into the action (\ref{eq:slicedAdS}), the on-shell bulk action can be written as
\begin{equation}
\vspace{1mm}
I [\hat{\Phi}] = - \frac{1}{2} \int d^4 p~\hat{\Phi} (p) \Sigma (p) \hat{\Phi} (-p), 
\vspace{1mm}
\end{equation}
where $\Sigma (p) = e^{-3ky_0} p \frac{F (p, y_0)}{G (p, y_0)}$ and
\begin{align}
\vspace{1mm}
G &\equiv J_{\alpha} \left( \frac{ip}{k_y} \right) Y_{\alpha \pm 1} \left( \frac{ip}{k_{y_1}} \right) - Y_{\alpha} \left( \frac{ip}{k_y} \right) J_{\alpha \pm 1} \left( \frac{ip}{k_{y_1}} \right) \nonumber \\
F &\equiv J_{\alpha \pm 1} \left( \frac{ip}{k_y} \right) Y_{\alpha \pm 1} \left( \frac{ip}{k_{y_1}} \right) - Y_{\alpha \pm 1} \left( \frac{ip}{k_y} \right) J_{\alpha \pm 1} \left( \frac{ip}{k_{y_1}} \right). \
\vspace{1mm}
\end{align}
Another useful quantity that can be computed in the five-dimensional theory is the conjugate variable $\check{\Phi} = - \delta I [\hat{\Phi}] ~/~ \delta \hat{\Phi}$. The original AdS/CFT correspondence recipe lets us determine the n-point functions $\langle \mathcal{O} ... \mathcal{O} \rangle$, for the gauge theory from the five-dimensional theory \cite{Maldacena:1997re,Witten:1998qj,Aharony:1999ti}. However, for general deformations of the bulk action $\int d^4p~\left( W [\check{\Phi}] + \varphi_0 \check{\Phi} / g_5 \right)$, where $g_5$ is an expansion parameter with dim$\left[ g_5 \right] = -1/2$, $W [\check{\Phi}]$ is an arbitrary function of $\check{\Phi}$, and $\varphi_0$ is the four-dimensional source, we follow Ref. \cite{Mueck:2002gm} to compute the improved correspondence formula. Let us call it \textit{dictionary I} and review here briefly. The Legendre transform of $I [\hat{\Phi}]$,
\begin{equation}
\vspace{1mm}
J [\check{\Phi}] = I - \int d^4p~ \hat{\Phi} \frac{\delta I}{\delta \hat{\Phi}},
\vspace{1mm}
\end{equation}
can be used to construct a generating functional $\mathcal{S}_{\text{holo}}$, from which one can compute the mass spectrum. Reference \cite{Mueck:2002gm} constructs the generating functional
\begin{equation}
\vspace{1mm}
\mathcal{S}_{\text{holo}} = J [\check{\Phi}] + \int d^4p~\left( W [\check{\Phi}] + \frac{1}{g_5} \varphi_0 \check{\Phi} \right),
\vspace{1mm}
\label{eq:holo}
\end{equation}
the minimization $\delta \mathcal{S}_{\text{holo}} / \delta \check{\Phi} = 0$ of which determines the relationship between $\check{\Phi}$ and the source $\varphi_0$. Plugging the solution for $\check{\Phi}$ back into Eq.(\ref{eq:holo}) results in a functional $\mathcal{S}_{\text{holo}} [\varphi_0]$. The AdS/CFT correspondence  can therefore be expressed as the following relation between the four- and five-dimensional theories,
\begin{equation}
\vspace{1mm}
e^{- \left( \mathcal{S}_{\text{holo}} [\varphi_0] - \mathcal{S}_{\text{holo}} [\varphi_0 =0] \right)} = \big\langle e^{- \int d^4p~\frac{1}{\Lambda_{\text{UV}}^{\Delta-3}} \varphi_0 \mathcal{O}} \big\rangle_{W[\mathcal{O}]},
\vspace{1mm}
\label{eq:AdSCFT}
\end{equation}
where $\Lambda_{\text{UV}} \equiv 2k_{y_0}$ is the cutoff scale. This definition can be considered as choosing an origin for the location of the UV brane in the fifth dimension. With this definition, following the literature and picking $y_0 = 0$, the curvature would obey the inequality $k / \Lambda_{\text{UV}} \lesssim 2$, in order for the classical metric solution to be valid \cite{Agashe:2007zd}. 

Let us start with no deformation of the gauge theory other than the addition of the source term, $W = 0$. For such linear deformations of the theory, i.e., single-trace deformations, the result of minimizing $\mathcal{S}_{\text{holo}}$, $\hat{\Phi} = - \varphi_0 / g_5$, is compatible with the original AdS/CFT recipe. This results in a trivial generating functional $\mathcal{S}_{\text{holo}} [\varphi_0] = I [ - \varphi_0 / g_5]$. Using the generating functional from the five-dimensional theory, and noting that $\mathcal{S}_{\text{holo}} \left[ \varphi_0 = 0 \right] = 0$ for $W=0$, we can compute the two-point functions
\begin{align}
\vspace{1mm}
\Lambda_{\text{UV}}^{2\Delta -6} \frac{\delta^2 \mathcal{S}_{\text{holo}} \left[ \varphi_0 \right]}{\delta \varphi_0 (p) \delta \varphi_0 (-p)} \Bigg|_{\varphi_0 = 0}&= \langle \mathcal{O} (p) \mathcal{O} (-p) \rangle + ... \nonumber \\
- \frac{\Lambda_{\text{UV}}^{2\Delta -6}}{2g_5^2} \Sigma (p) &\sim \begin{cases} p^{2\alpha} + ... & y_1 \rightarrow \infty \\ \sum_n \frac{a_n^2}{p^2 + m_n^2} & \text{finite}~y_1, \\ \end{cases}
\vspace{1mm}
\label{eq:twopoint}
\end{align}
where $a_n = \langle 0 | \mathcal{O} | \varphi^n \rangle$ is the matrix element for $\mathcal{O}$ to create the $n$th composite state $\varphi^n$, with mass $m_n$ from the vacuum \cite{tHooft:2002ufq,Witten:1979kh}, and ... denotes the analytical or diverging terms that we do not care about while calculating the scaling dimension of $\mathcal{O}$. For the case $y_1 \rightarrow \infty$, we only showed the leading nonanalytical term that gives us the dimension $\Delta = 2 + \alpha$. The mass spectrum of the composite states after confinement can then be found by calculating the poles of $\Sigma (p)$ for finite $y_1$, which are given by the solutions of the following equation:
\begin{equation}
\vspace{1mm}
\frac{J_{\alpha} \left( m_n / k_{y_0} \right)}{Y_{\alpha} \left( m_n / k_{y_0} \right)} = \frac{J_{\alpha \pm 1} \left( m_n / k_{y_1} \right)}{Y_{\alpha \pm 1} \left( m_n / k_{y_1} \right)}.
\vspace{1mm}
\label{eq:CFT}
\end{equation}
This means that the bulk field can be decomposed as a tower of the composite states
\begin{equation}
\vspace{1mm}
\Phi (x, y) = \sum_{n=1}^{\infty} \varphi^n (x) g^n (y).
\vspace{1mm}
\end{equation}
The profiles $g^n (y)$ that satisfy the bulk equation of motion (\ref{eq:eom}), with $\Box \varphi^n = m_n^2 \varphi^n$, can be written as
\begin{equation}
\vspace{1mm}
g^n (y) = N_n e^{2ky} \left[ J_{\alpha} \left( \frac{m_n}{k_y} \right) + \kappa (m_n) Y_{\alpha} \left( \frac{m_n}{k_y} \right) \right],
\vspace{1mm}
\end{equation}
where $N_n$ is the normalization constant and $\kappa (m_n)$ is determined by imposing the boundary conditions. Consider the lightest mode in the mass spectrum coming from Eq.(\ref{eq:CFT}), with mass
\begin{equation}
\vspace{1mm}
m_1 \sim \begin{cases} k_{y_1} & -~\text{branch} \\ k_{y_1} e^{\alpha k(y_0 - y_1)} & +~\text{branch}. \\ \end{cases}
\vspace{1mm}
\label{eq:m1}
\end{equation}
Note that only the $+$ branch has a mode whose mass is lighter than the confinement scale $\Lambda_{\text{IR}} \equiv 2k_{y_1}$ since $\alpha > 0$ and $y_0 < y_1$. To compute the mass spectrum for the pure composite states, one needs to remove the UV brane, $y_0 \rightarrow -\infty$. This tells us that the $+$ branch had a massless mode but it is modified by the finite UV cutoff effects. On the other hand, the $-$ branch never has a mode lighter than the confinement scale. Equation (\ref{eq:CFT}) is equivalent to the following boundary conditions for the profiles of the composite states:
\begin{equation}
\vspace{1mm}
g^n (y_0) = 0 \quad \text{and} \quad \left[ \left( \partial_y - bk \right) g^n (y) \right]_{y = y_1} = 0.
\vspace{1mm}
\end{equation}
Note that this implies a fixed UV boundary value of the bulk field, $\left[ \delta \Phi \right]_{y = y_0} = 0$, even though the composite states are dynamical fields with $\delta \varphi^n \neq 0$. Since $\varphi_0$ is not a dynamical field, one needs to set it to zero at the end of the calculations to be consistent with the boundary conditions of the profiles. This calculation lets us conclude that according to dictionary I, the action (\ref{eq:slicedAdS}) with
\begin{equation}
\vspace{1mm}
\left[ \delta \Phi \right]_{y = y_0} = 0
\vspace{1mm}
\end{equation}
is the five-dimensional dual of a four-dimensional gauge theory with the operator $\mathcal{O}$ and $m_1$ is the smallest mass of the composite spectrum created by this operator. Therefore, we need to modify the way we approach the five-dimensional action to have a massless or a light mode in the four-dimensional theory, especially for the $-$ branch. Note that the composite states' nature might change after these modifications. Now, we consider different ways of modifying the theory for that purpose.

\subsection{Dictionary I: $\left[ \delta \Phi \right]_{y = y_0} \neq 0 \Rightarrow \varphi^s, \mathcal{O}$}
\label{subsec:linear mixing}

The approach to have a massless mode has been to work with a dynamical UV boundary value instead, $\left[ \delta \Phi \right]_{y = y_0} \neq 0$, which means that the UV boundary condition is modified to be
\begin{equation}
\vspace{1mm}
\left[ \left( \partial_y - bk \right) \Phi ~ \right]_{y=y_0} = 0.
\vspace{1mm}
\end{equation}
According to dictionary I, this means that in the four-dimensional dual theory we introduce an elementary scalar field $\varphi^s$ that linearly mixes with the composite operator $\mathcal{O}$. This is also known as the single-trace deformation of the theory, $\int d^4 p~\varphi^s \mathcal{O} / \Lambda_{\text{UV}}^{\Delta-3}$, which means $W[\mathcal{O}] = 0$. Then, the five-dimensional theory is deformed by $\int d^4p~\varphi^s \check{\Phi} / g_5$, and observe that this is very similar to the term that we added to compute the two-point functions. The difference is that now $\varphi^s$ is a dynamical field with $\delta \varphi^s \neq 0$ unlike the fixed source $\varphi_0$. $\varphi^s$ mixes with the composite states, and the mass spectrum of the theory is modified. The modified mass spectrum can be computed by minimizing the effective action $\mathcal{S}_{\text{holo}} [\varphi^s]$. Minimizing the effective action, $\delta \mathcal{S}_{\text{holo}} [\varphi^s] = \frac{\delta \mathcal{S}_{\text{holo}} [\varphi^s]}{\delta \varphi^s (p)} \delta \varphi^s = 0$, requires $\Sigma (p)$ to be zero since $\varphi^s (-p) \neq 0$, which is satisfied only for certain momentum values. Therefore, the modified mass spectrum $M_n$ is given by the zeros of $\Sigma (p)$ instead of its poles. The zeros are given by 
\begin{equation}
\vspace{1mm}
\frac{J_{\alpha \pm 1} \left( M_n / k_{y_0} \right)}{Y_{\alpha \pm 1} \left( M_n / k_{y_0} \right)} = \frac{J_{\alpha \pm 1} \left( M_n / k_{y_1} \right)}{Y_{\alpha \pm 1} \left( M_n / k_{y_1} \right)}.
\vspace{1mm}
\label{eq:KK}
\end{equation}
In this case, there is a massless eigenstate, $M_0 = 0$. We conclude that these modified mass eigenstates are admixtures of $\varphi^s$ and $\varphi^n$. Then, the decomposition of the bulk field should be supplemented by a new four-dimensional field $\varphi^s (x)$, with a profile $g^s (y)$ that has a nonzero value on the UV brane, 
\begin{equation}
\vspace{1mm}
\Phi (x,y) = \varphi^s (x) g^s (y) + \sum_{n=1}^{\infty} \varphi^n (x) g^n (y),
\vspace{1mm}
\end{equation} 
where $g^s (y)$ also satisfies the bulk equation of motion (\ref{eq:eom}), with $\Box \varphi^s = m_s^2 \varphi^s$, and it is normalized so that the kinetic terms in the resulting four-dimensional theory are canonical. This is called the holographic basis. The new profile $g^s (y)$ can also be written as
\begin{equation}
\vspace{1mm}
g^s (y) = N_s e^{2ky} \left[ J_{\alpha} \left( \frac{m_s}{k_y} \right) + \kappa (m_s) Y_{\alpha} \left( \frac{m_s}{k_y} \right) \right],
\vspace{1mm}
\end{equation}
where $N_s$ is the normalization constant and $m_s$ is the mass term for $\varphi^s$. The constant $\kappa (m_s)$ is determined by imposing the boundary conditions. Since $\varphi^s$ is proportional to $\Phi (p, y_0)$, we expect its profile to satisfy the boundary condition
\begin{equation}
\vspace{1mm}
\left \{ \left[ \partial_y - (2 - \alpha)k \right] g^s (y) \right \}_{y = y_0} = 0,
\vspace{1mm}
\label{eq:UVcond}
\end{equation}
from Eq.(\ref{eq:nearUV}). This boundary condition makes sure that the profile $g^s (y)$ looks like the dominant term in Eq.(\ref{eq:nearUV}) near the UV boundary. On the other hand, different boundary conditions for $g^s (y)$ on the IR brane can be imposed, which in turn determines the nature of the mixing between the elementary and composite scalars.

\subsubsection{\textbf{IR condition I}}
\label{subsubsec:traditional}

\noindent For example, Ref. \cite{Batell:2007jv} implicitly picks 
\begin{equation}
\vspace{1mm}
\left \{ \left[ \partial_y - (2 - \alpha)k \right] g^s (y) \right \}_{y = y_1} = 0,
\vspace{1mm}
\label{eq:condition1}
\end{equation}
which brings the following equation for the $-$ branch,
\begin{equation}
\vspace{1mm}
\frac{J_{\alpha - 1} \left( m_s / k_{y_0} \right)}{Y_{\alpha - 1} \left( m_s / k_{y_0} \right)} = \frac{J_{\alpha - 1} \left( m_s / k_{y_1} \right)}{Y_{\alpha - 1} \left( m_s / k_{y_1} \right)},
\vspace{1mm}
\end{equation}
which is the same as the $-$ branch in Eq.(\ref{eq:KK}). Therefore, one of the solutions to this equation is $m_s = 0$. In this case, there is only a kinetic mixing between $\varphi^s$ and $\varphi^n$ in the resulting four-dimensional theory. However, for the $+$ branch, Ref. \cite{Batell:2007jv} finds both kinetic and mass mixing. Diagonalizing the kinetic and mass matrices, we can see that there is a massless eigenstate $\phi_0$. The main conclusion for this way of achieving a massless mode is that the massless scalar is an admixture of elementary and composite scalars; for example, in the $-$ branch, we have
\begin{equation}
\vspace{1mm}
\phi_0 \approx \mathcal{N}_0 \left( \varphi^s + \varepsilon_1 \varphi^1 + ... \right),~\text{where}~\varepsilon_1 \sim \sqrt{\frac{2(b-1)}{1-e^{2(1-b)\pi kR}}} 
\vspace{1mm}
\label{eq:masslessmode}
\end{equation}
and $\mathcal{N}_0$ is the normalization constant. If we consider the dominant contribution, the massless eigenstate
\begin{equation}
\vspace{1mm}
\phi_0 ~ \text{is mostly} ~ \begin{cases} \text{elementary}~ \varphi^s, & - ~ \text{branch} \\ \text{composite} ~ \varphi^n, & + ~ \text{branch}. \\ \end{cases}
\vspace{1mm}
\end{equation}

\subsubsection{\textbf{IR condition II}}
\label{subsubsec:alternative}

\noindent Let us pick 
\begin{equation}
\vspace{1mm}
g^s (y_1) = 0,
\vspace{1mm}
\end{equation}
instead of the modified Neumann condition that we used before, Eq.(\ref{eq:condition1}). Combined with the usual boundary condition on the UV brane, Eq.(\ref{eq:UVcond}), it leads to the following equation for $b < 2$,
\begin{equation}
\vspace{1mm}
\frac{J_{\alpha -1} \left( m_s / k_{y_0} \right)}{Y_{\alpha -1} \left( m_s / k_{y_0} \right)} = \frac{J_{\alpha} \left( m_s / k_{y_1} \right)}{Y_{\alpha} \left( m_s / k_{y_1} \right)},
\vspace{1mm}
\end{equation}
that can be solved for new nonzero $m_s$ values. The smallest solution is given by
\begin{equation}
\vspace{1mm}
m_s \sim \begin{cases} k_{y_1} e^{(\alpha -1)k(y_0 - y_1)} & \alpha > 1 \\ k_{y_1} & 0 \leq \alpha < 1. \\ \end{cases}
\vspace{1mm}
\label{eq:sourcemass}
\end{equation}
We showed that there is no $m_s = 0$ for the $-$ branch unlike the case in Sec.\ref{subsubsec:traditional}. Now, we need to check whether this new solution also allows the system to have a massless eigenstate. Inserting the holographic basis into the action, we find
\begin{equation}
\vspace{1mm}
\mathcal{S_{\text{mix}}} = \int d^4 x \sum_{n=1}^{\infty} \left[ - \frac{1}{2} z_n \partial_{\mu} \varphi^s \partial^{\mu} \varphi^n - \frac{1}{2} \mu_n^2 \varphi^s \varphi^n \right],
\vspace{1mm}
\end{equation}
where $\mathcal{S_{\text{mix}}}$ includes the kinetic $z_n$ and mass mixings $\mu_n^2$. 
If we can show that the determinant of the mass matrix is zero, we are done with proving that there is a massless eigenstate. For this purpose, we need to know what the mass mixing terms $\mu_n^2$ are. Using the profiles and integrating by parts in two different ways, they are given by
\begin{align}
\vspace{1mm}
\mu_n^{2A} &\equiv \int_{y_0}^{y_1} dy~m_s^2 g^s \left[ e^{-2ky} g^n - g^n (y_1) e^{- bky_1} e^{(b-2)ky} \right] \nonumber \\ \mu_n^{2B} &\equiv \int_{y_0}^{y_1} dy~m_n^2 g^n \left[ e^{-2ky} g^s - g^s (y_0) e^{-bky_0} e^{(b-2)ky} \right], \
\vspace{1mm}
\end{align}
where $\mu_n^2 = \mu_n^{2A} = \mu_n^{2B}$. The smallest eigenvalue of the mass matrix is zero if 
\begin{equation}
\vspace{1mm}
\sum_{n=1}^{\infty} \frac{ \mu_n^4}{m_n^2} = m_s^2.
\vspace{1mm}
\end{equation}
This equation can be shown to hold by computing $\sum_{n=1}^{\infty} \mu_n^{2A} \mu_n^{2B} / m_n^2$ with the help of the completeness relation
\begin{equation}
\vspace{1mm}
\sum_{n=1}^{\infty} g^n (y) g^n (y^{\prime}) = e^{2ky} \delta (y - y^{\prime}).
\vspace{1mm}
\end{equation}
Therefore, there is a massless eigenstate in this holographic basis as well. The difference from the holographic basis in Sec.\ref{subsubsec:traditional} is that there are both kinetic and mass mixings in this basis for both branches. A truncated version of this type of mass mixing is observed in a supersymmetric model \cite{Buyukdag:2018cka}. Note that the results of Ref. \cite{Buyukdag:2018cka} come from a purely four-dimensional consideration. Again, the main conclusion for this way of achieving a massless eigenstate is that the massless scalar is an admixture of elementary and composite scalars; for example, in the $-$ branch, it is given by Eq.(\ref{eq:masslessmode}). If we consider the dominant contribution, the massless eigenstate
\begin{equation}
\vspace{1mm}
\phi_0 ~ \text{is mostly} ~ \begin{cases} \text{elementary}~  \varphi^s, & - ~ \text{branch} \\ \text{composite}~ \varphi^n, & + ~ \text{branch}. \\ \end{cases}
\vspace{1mm}
\end{equation}
This calculation lets us conclude that according to dictionary I, the action (\ref{eq:slicedAdS}) with
\begin{equation}
\vspace{1mm}
\left[ \delta \Phi \right]_{y = y_0} \neq 0
\vspace{1mm}
\end{equation}
is the five-dimensional dual of a four-dimensional gauge theory with the elementary field $\varphi^s$ and the composite operator $\mathcal{O}$. There is a partially composite massless eigenstate in the spectrum.

\subsection{Dictionary II: $\left[ \delta \Phi \right]_{y = y_0} \neq 0 \Rightarrow \mathcal{O}^{\prime}$}
\label{subsec:multi trace}

Here, the UV boundary condition is modified to be
\begin{equation}
\vspace{1mm}
\left[ \left( \partial_y - bk \right) \Phi ~ \right]_{y=y_0} = 0
\vspace{1mm}
\label{eq:UVboundary}
\end{equation}
as well. According to \textit{dictionary II}, we construct the generating functional $\mathcal{S}_{\text{holo}}^{\prime}$ in the five-dimensional theory differently,
\begin{equation}
\vspace{1mm}
\mathcal{S}^{\prime}_{\text{holo}} = I [ \hat{\Phi} ] + \int d^4p~\frac{\Lambda_{\text{UV}}}{g_5} \varphi_0 \hat{\Phi},
\vspace{1mm}
\label{eq:modifiedgenerating}
\end{equation}
the minimization of which, $\delta \mathcal{S}_{\text{holo}}^{\prime} / \delta \hat{\Phi} = 0$, determines the relationship between $\hat{\Phi}$ and the source $\varphi_0$. This is an example of the interpretation in Ref. \cite{Klebanov:1999tb} where the roles of two solutions near the AdS boundary, $\hat{\Phi}$ and $\check{\Phi}$, are interchanged. Plugging the solution for $\hat{\Phi}$ back into Eq.(\ref{eq:modifiedgenerating}) results in a functional
\begin{equation}
\vspace{1mm}
\mathcal{S}_{\text{holo}}^{\prime} [\varphi_0] = - \int d^4p~\frac{\Lambda_{\text{UV}}^2}{2g_5^2} \varphi_0 (p) \frac{1}{\Sigma (p)} \varphi_0 (-p).
\vspace{1mm}
\end{equation}
Using this generating functional, we can compute the two point functions,
\begin{align}
\vspace{1mm}
\frac{1}{\Lambda_{\text{UV}}^{2+2\alpha}} \frac{\delta^2 \mathcal{S}_{\text{holo}}^{\prime} \left[ \varphi_0 \right]}{\delta \varphi_0 (p) \delta \varphi_0 (-p)} \Bigg|_{\varphi_0 = 0} &= \langle \mathcal{O}^{\prime} (p) \mathcal{O}^{\prime} (-p) \rangle + ... \nonumber \\
- \frac{1}{2g_5^2\Lambda_{\text{UV}}^{2\alpha}} \frac{1}{\Sigma (p)} &\sim \begin{cases} p^{-2\alpha} + ... & y_1 \rightarrow \infty \\ \sum_n \frac{a_n^2}{p^2 + m_n^2} & \text{finite}~y_1, \\ \end{cases}
\vspace{1mm}
\label{eq:Oprime}
\end{align}
where $\mathcal{O}^{\prime}$ is the composite operator in the four-dimensional dual theory according to dictionary II. For the case $y_1 \rightarrow \infty$, we only showed the leading nonanalytical term that gives us the dimension $\Delta^{\prime} = 2 - \alpha$. This interpretation should then be restricted to the parameter values $0 \leq \alpha \leq 1$ so that $\Delta^{\prime} \geq 1$. The mass spectrum of the composite states after confinement can be found by calculating the zeros of $\Sigma (p)$ for finite $y_1$, which are given by the solutions of the following equation:
\begin{equation}
\vspace{1mm}
\frac{J_{\alpha \pm 1} \left( M_n / k_{y_0} \right)}{Y_{\alpha \pm 1} \left( M_n / k_{y_0} \right)} = \frac{J_{\alpha \pm 1} \left( M_n / k_{y_1} \right)}{Y_{\alpha \pm 1} \left( M_n / k_{y_1} \right)}.
\vspace{1mm}
\label{eq:massless}
\end{equation}
Note that this is the same mass spectrum as in Eq.(\ref{eq:KK}). However, according to this dictionary, the massless eigenstate is not an admixture of $\varphi^s$ and $\varphi^n$ as in Sec.\ref{subsec:linear mixing}. It is the pure composite state $\varphi^{\prime 0}$ created by $\mathcal{O}^{\prime}$ at the confinement scale.

Now, let us write the profiles for the composite states with mass values calculated from Eq.(\ref{eq:massless}). The bulk field can be decomposed as a tower of the composite states,
\begin{equation}
\vspace{1mm}
\Phi (x, y) = \sum_{n=0}^{\infty} \varphi^{\prime n} (x) f^n (y),
\vspace{1mm}
\end{equation}
where $\varphi^{\prime n}$ are the new composite states and $f^n (y)$ are the new profiles. These profiles $f^n (y)$ that satisfy the bulk equation of motion (\ref{eq:eom}), with $\Box \varphi^{\prime n} = M_n^2 \varphi^{\prime n}$, can be written as 
\begin{equation}
\vspace{1mm}
f^n (y) = N_n e^{2ky} \left[ J_{\alpha} \left( \frac{M_n}{k_y} \right) + \kappa (M_n) Y_{\alpha} \left( \frac{M_n}{k_y} \right) \right],
\vspace{1mm}
\end{equation}
where $N_n$ is the normalization constant and $\kappa (M_n)$ is determined by imposing the boundary conditions. Equation (\ref{eq:massless}) is equivalent to the following boundary conditions for the profiles,
\begin{equation}
\vspace{1mm}
\left[ \left( \partial_y - bk \right) f^n \right]_{y = y_0,y_1} = 0,
\vspace{1mm}
\end{equation}
which is consistent with Eq.(\ref{eq:UVboundary}). Note that so far we did not need to consider different branches separately. Since $0 \leq \alpha \leq 1$ in this dictionary, the boundary mass parameter can take values in the range $1 \leq b \leq 3$. If we look at the profile of the massless mode with respect to the flat metric \cite{Gherghetta:2010cj},
\begin{equation}
\begin{split}
\vspace{1mm}
\tilde{f}^0 (y) &\equiv e^{-ky} f^0 (y) \\
&\propto e^{(b-1)ky}, \\
\end{split}
\end{equation}
we see that the profile for all possible $b$ values is localized toward the IR brane. The fact that the IR brane is associated with the composite sector of the four-dimensional dual theory is consistent with our conclusion that says the massless mode
\begin{equation}
\vspace{1mm}
\phi_0 ~ \text{is the pure composite state} ~ \varphi^{\prime 0} ~ \text{for both branches} \nonumber
\vspace{1mm}
\end{equation} 
according to dictionary II unlike the partially composite massless eigenstate in dictionary I. This calculation lets us conclude that according to dictionary II the action (\ref{eq:slicedAdS}) with
\begin{equation}
\vspace{1mm}
\left[ \delta \Phi \right]_{y = y_0} \neq 0
\vspace{1mm}
\end{equation}
is the five-dimensional dual of a four-dimensional gauge theory with the operator $\mathcal{O}^{\prime}$ and there is a composite massless mode in the spectrum created by this operator. The results are summarized in the Table \ref{tab:scalingtable}.

\begin{table*}[t]
  \centering
  \begin{minipage}{1.0\textwidth}
    \centering
    \resizebox{\columnwidth}{!}{
    \begin{tabular}{c c c c c}
      \hline \hline
      Dictionary & Fields & Massless eigenstate & \multicolumn{2}{c}{Boundary conditions} \\
      \hline

      I & \makecell{$\varphi^s, \mathcal{O}$ \\ \rule{0pt}{2ex} $\Delta = 2 + \alpha$ \\ $0 \leq \alpha$ \\} & \makecell{$-$ branch: mostly elementary $\varphi^s$ \\  $+$ branch: mostly composite $\varphi^n$ \\ }&   \multicolumn{2}{c}{\makecell{ \rule{0pt}{4ex} $\left \{ \left[ \partial_y - (2 - \alpha)k \right] g^s (y) \right \}_{y = y_0} = 0$ \\ \rule{0pt}{4ex} $g^n (y_0) = 0$ \\ \rule{0pt}{4ex} $\left[ \left( \partial_y - bk \right) g^n (y) \right]_{y = y_1} = 0$ \\  \\   } } \\

     & &  & \makecell{ \rule{0pt}{4ex} IR condition I \\ \hline $\left \{ \left[ \partial_y - (2 - \alpha)k \right] g^s (y) \right \}_{y = y_1} = 0$ \rule{0pt}{4ex}  \\ \\ } &  \makecell{\rule{0pt}{4ex} IR condition II \\ \hline $g^s (y_1) = 0$ \rule{0pt}{4ex} \\ \\} \\

      II & \makecell{ \\ $\mathcal{O}^{\prime}$ \\ \rule{0pt}{2ex} $\Delta^{\prime} = 2 - \alpha$ \\ $0 \leq \alpha \leq 1$ \\ \\ }& $\pm$ branch: pure composite $\varphi^{\prime 0}$ & \multicolumn{2}{c}{ $\left[ \left( \partial_y - bk \right) f^n (y) \right]_{y = y_0,y_1} = 0$  \rule{0pt}{4ex} } \\ 
      
      \hline \hline  
    \end{tabular}}
  \end{minipage}
  \caption{The summary of the four- and five-dimensional results for both dictionaries. The boundary mass parameter can take values $-\infty < b < \infty$ in dictionary I and values $1 \leq b \leq 3$ in dictionary II. 
    \label{tab:scalingtable}}
\end{table*}

\section{Relationship between $\mathcal{O}$ and $\mathcal{O}^{\prime}$}
\label{sec:relationship}

This result $\Delta^{\prime} = 2 - \alpha$ is very similar to the one in Ref. \cite{Klebanov:1999tb}, where the Legendre transform of the on-shell bulk action inverts the expression for the two-point function $\langle \mathcal{O} (p) \mathcal{O} (-p) \rangle$. Let us try to explain the relationship between two dictionaries, which is an attempt to show how $\mathcal{O}$ and $\mathcal{O}^{\prime}$ are related. Remember that the smallest mass in the spectrum created by $\mathcal{O}$ is $m_1$ given by Eq.(\ref{eq:m1}) while the smallest mass in the spectrum created by $\mathcal{O}^{\prime}$ is $M_0 = 0$.

\begin{figure}[t]
  \includegraphics[width=\columnwidth]{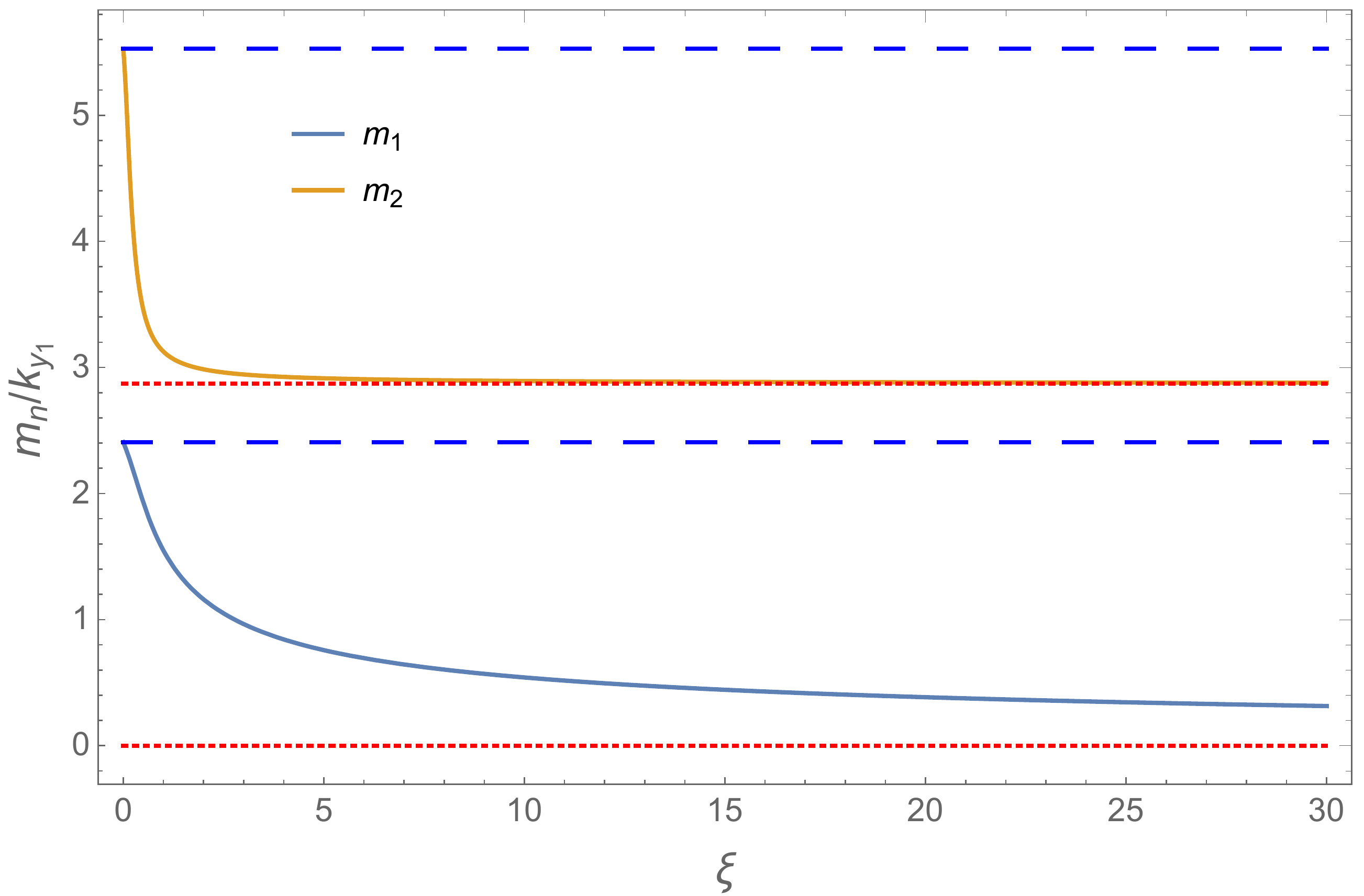}
  \caption{The lightest and the second-lightest modes of the spectrum as a function of the deformation, $\xi$, for $e^{ky_0} = 0.1$, $g_5^2\Lambda_{\text{UV}} = 1$, $\alpha = 1$, and $\pi k (y_1 - y_0) = 4$. Horizontal dashed lines show the poles of $\Sigma (p)$, while the horizontal dotted lines show the zeros of $\Sigma (p)$.}
  \label{fig:multitracefig}
\end{figure}

We introduce a new interaction $W[\mathcal{O}] = - \xi \mathcal{O}^2 / \Lambda_{\text{UV}}^{2\Delta -4}$ into the initial model with no massless mode (dictionary I with $\left[ \delta \Phi \right]_{y = y_0} = 0$), where $\xi$ is a dimensionless constant. This is also known as the multitrace deformation of the gauge theory. Since we did not introduce a dynamical field like $\varphi^s$ in this case, we need to add an additional deformation $\varphi_0 \mathcal{O} / \Lambda_{\text{UV}}^{\Delta -3}$, where $\delta \varphi_0 = 0$ as before to be able to probe the theory and calculate the mass spectrum. To determine the five-dimensional counterpart of this deformation, we need to know the relationship between $\check{\Phi}$ and $\langle \mathcal{O} \rangle$. This can be achieved by comparing the linear deformations with $\varphi_0$ from both sides, 
\begin{equation}
\vspace{1mm}
\frac{\check{\Phi}}{g_5} \leftrightarrow \frac{\langle \mathcal{O} \rangle}{\Lambda_{\text{UV}}^{\Delta -3}}.
\vspace{1mm}
\end{equation}
Then, the five-dimensional theory is deformed by $\int d^4p\left( - \xi \check{\Phi}^2 / (g_5^2 \Lambda_{\text{UV}}^2) + \varphi_0 \check{\Phi} / g_5 \right)$. For such multitrace deformations of the theory, minimizing $\mathcal{S}_{\text{holo}}$ results in a more complicated generating functional \cite{Mueck:2002gm}
\begin{equation}
\vspace{1mm}
\mathcal{S}_{\text{holo}} \left[ \varphi_0 \right] = - \frac{1}{2} \int d^4p~\varphi_0 (p) \frac{\Sigma (p) / g_5^2}{1 - \xi \Sigma (p) / (g_5^2 \Lambda_{\text{UV}}^2)} \varphi_0 (-p).
\vspace{1mm}
\label{eq:genfunc}
\end{equation}
Using the generating functional from the five-dimensional theory and noting that $\mathcal{S}_{\text{holo}} \left[ \varphi_0 = 0 \right] = 0$, we can compute the two-point functions,
\begin{align}
\vspace{1mm}
\Lambda_{\text{UV}}^{2\Delta -6} \frac{\delta^2 \mathcal{S}_{\text{holo}} \left[ \varphi_0 \right]}{\delta \varphi_0 (p) \delta \varphi_0 (-p)} \Bigg|_{\varphi_0 = 0} &= \langle \mathcal{O} (p) \mathcal{O} (-p) \rangle + ... \nonumber \\
-\frac{\Lambda_{\text{UV}}^{2\Delta -4}}{2} \frac{\Sigma (p)}{g_5^2 \Lambda_{\text{UV}}^2 - \xi \Sigma (p)} &\sim \sum_n \frac{a_n^2}{p^2 + m_n^2} ~ \text{for finite}~y_1. \
\vspace{1mm}
\label{eq:multitrace}
\end{align}
The expression for $y_1 \rightarrow \infty$ would not be as trivial as the one in Eq.(\ref{eq:twopoint}). This means that the dimension of the operator is not trivially related to the bulk mass parameter anymore. The mass spectrum of the composite states after confinement are then given by the solutions of the following equation:
\begin{equation}
\vspace{1mm}
\frac{1}{\xi} = \frac{\Sigma (p)}{g_5^2 \Lambda_{\text{UV}}^2}.
\vspace{1mm}
\end{equation}
This equality makes the denominator in Eq.(\ref{eq:multitrace}) zero while keeping the numerator nonzero. Note that the mass spectrum of the original gauge theory, $\xi = 0$, was given by the poles of $\Sigma (p)$. If we consider the limit $\xi \rightarrow \infty$, the solutions $m_n$ are given by the zeros of $\Sigma (p)$. The mass spectrum of the maximally deformed theory, $\xi \rightarrow \infty$, is then computed by using Eq.(\ref{eq:KK}). Therefore, this particular type of maximal deformation modifies the mass spectrum such that $m_n = M_n$ and there is a massless scalar without introducing a new elementary field like $\varphi^s$. How the mass eigenvalues change from the poles of $\Sigma (p)$ to the zeros of $\Sigma (p)$ can be seen in Fig.{\ref{fig:multitracefig}}. 
This phenomenon is also observed in a string theory setup \cite{Abt:2019tas}, where a light composite fermion is achieved in a similar limit. 

In this process, how $\xi$ can take large values is not obvious, but we can understand its physical implications. Expanding Eq.(\ref{eq:multitrace}) in this limit,  
\begin{equation}
\vspace{1mm}
\lim_{\xi \rightarrow \infty} \frac{\Lambda_{\text{UV}}^{2\Delta -4}}{2} \frac{-\Sigma (p)}{g_5^2\Lambda_{\text{UV}}^2 - \xi \Sigma (p)} \sim - \frac{\Lambda_{UV}^{2\Delta -4}}{2\xi} + \frac{g_5^2\Lambda_{UV}^{2\Delta -2}}{2\xi^2} \frac{1}{\Sigma (p)}, 
\vspace{1mm}
\label{eq:maxdeform}
\end{equation}
and redefining the operator $\mathcal{O} \rightarrow \mathcal{O} / \xi$, we can focus on the leading nonanalytical term
\begin{equation}
\vspace{1mm}
\langle \mathcal{O} (p) \mathcal{O} (-p) \rangle \sim p^{-2\alpha} ~~ \text{for} ~~ y_1 \rightarrow \infty, 
\vspace{1mm}
\end{equation}
where we suppressed the dimensionful constants. Note that this expression looks just like Eq.(\ref{eq:Oprime}), which implies that the operator in dictionary II, $\mathcal{O}^{\prime}$ is the meaningful degree of freedom for the operator in dictionary I, $\mathcal{O}$ in this limit $\xi \rightarrow \infty$. Considering the AdS/CFT formula (\ref{eq:AdSCFT}), this result can be summarized as
\begin{equation}
\vspace{1mm}
\lim_{\xi \rightarrow \infty} \Big\langle e^{- \int d^4p~\frac{1}{\Lambda_{\text{UV}}^{\Delta-3}} \varphi_0 \mathcal{O}} \Big\rangle_{-\frac{\xi \mathcal{O}^2}{\Lambda_{\text{UV}}^{2\Delta -4}}} \rightarrow \Big\langle e^{- \int d^4p~\frac{1}{\Lambda_{\text{UV}}^{\Delta^{\prime}-3}} \varphi_0 \mathcal{O}^{\prime}} \Big\rangle_{0}.
\vspace{1mm}
\end{equation}
The same limit in the five-dimensional theory then is equivalent to transitioning,
\begin{equation}
\vspace{1mm}
\text{from}~~\left[ \delta \Phi \right]_{y=y_0} = 0 ~~ \text{to} ~~ \left[ \left( \partial_y - bk \right) \Phi \right]_{y=y_0} = 0,
\vspace{1mm}
\end{equation}
without adding an external elementary field $\varphi^s$.

\section{Conclusion}

We showed how a five-dimensional action can be used to describe two different strongly coupled gauge theory with a massless mode. For example, the five-dimensional parameter $\alpha$ is mapped to two different four-dimensional parameters, $\Delta$ and $\Delta^{\prime}$. An important observation was that an effective interaction term in the strong sector can modify the composite spectrum with no light state such that there is a pure composite scalar lighter than the confinement scale. How much lighter this state is depends on the interaction strength. The physical origin of this interaction strength and how it can take large values are subjects that need further investigation. However, when one does not need a big suppression from the confinement scale, a reasonable interaction strength is useful enough. 

A light composite scalar as we discussed in this paper might be used for dark matter models with light scalar mediators. Moreover, the strong dynamics that produce a composite Higgs boson could also give rise to such light composite scalars, which would affect the Higgs decay channels. Or more drastically, Higgs boson itself could be imagined as the light composite state of a strongly coupled gauge theory with a confinement scale around TeV. This kind of model would allow for heavier composite scalars to be above the several TeV range that is consistent with LHC results.

\begin{acknowledgments}

We thank Brian Batell and Tony Gherghetta for helpful discussions. This work is supported in part by the DOE Grant No. DE-SC0011842 at the University of Minnesota. 

\end{acknowledgments}

\appendix

\end{document}